\documentstyle[epsf]{article}
\begin{document}
\title{\textbf{$\eta^{\prime}$ production in proton-proton scattering 
close to threshold}}  
\author{}
\date{}
\maketitle
\begin{figure}[ht]
\leavevmode
   \epsfxsize=14.5cm
  \epsfysize=16.5cm
   \hspace{1.5cm}
   \epsffile{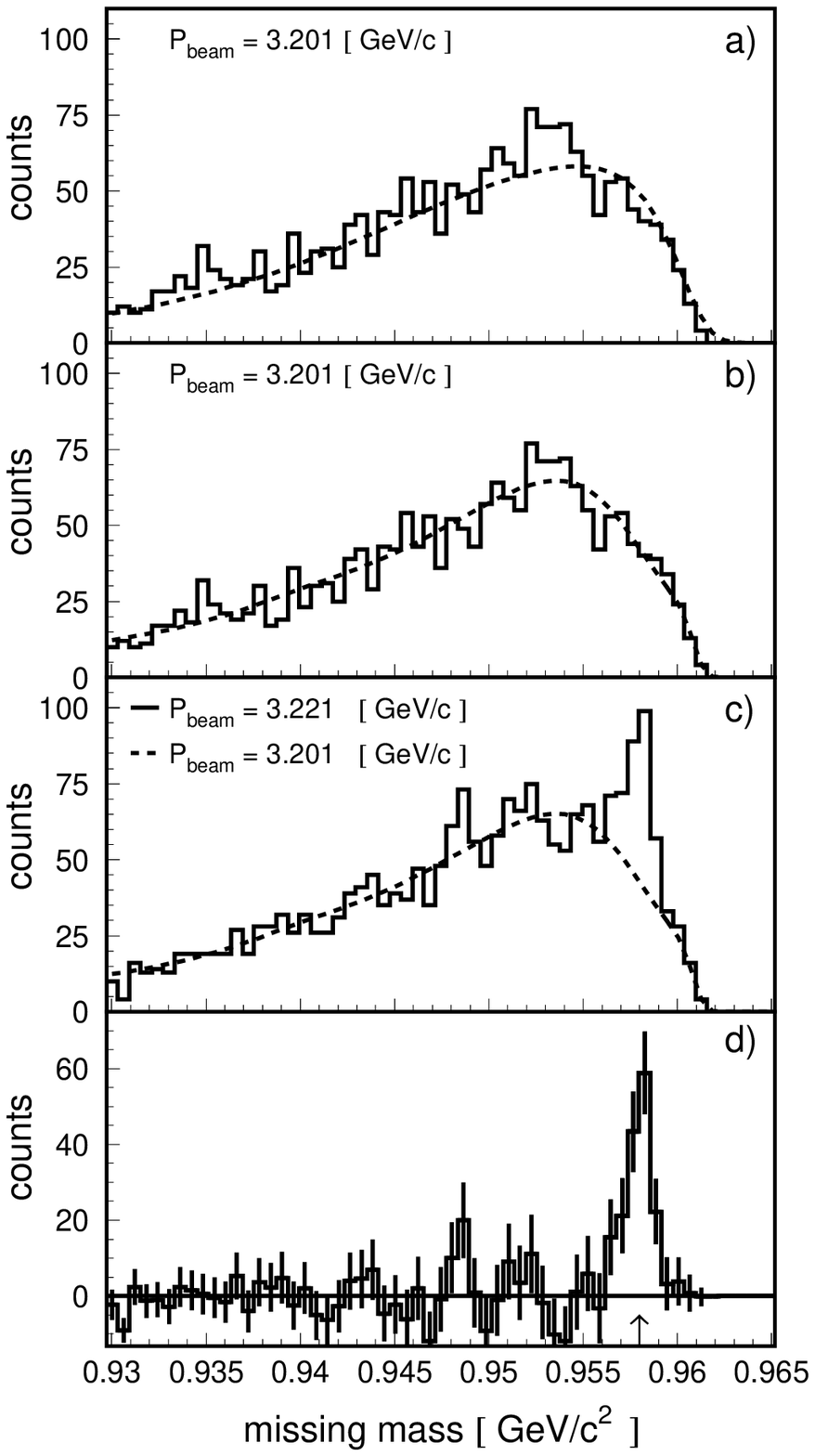}
\caption{
Missing mass spectra of the unobserved particle X in the reaction
$pp \rightarrow ppX $: \protect \\
a) data at a beam momentum below threshold (solid line) and  \protect \\
\mbox{~~~~~MC calculations for the reactions 
 $pp \rightarrow pp \pi^+ \pi^- $ and $pp \rightarrow pp \pi^+ \pi^- \pi^0 $ 
 (dashed line),} \protect \\ 
 b) data (solid line), smooth fit function to the data (dashed line), \protect \\
c) data at a beam momentum of 3.221 GeV/c for the $\eta^{\prime}$ production 
(solid line), 
\mbox{~~~~~scaled background from b) (dashed line)} \protect \\
d) difference between solid and dashed lines of c), the arrow indicates the $\eta^{\prime}$ 
mass. 
}
  \label{eta_prime_missing_mass}
\end{figure}
\begin{figure}[ht]
\leavevmode
 \epsfxsize=7.0cm
   \epsfysize=7.0cm
   \hspace{1cm}
    \epsffile{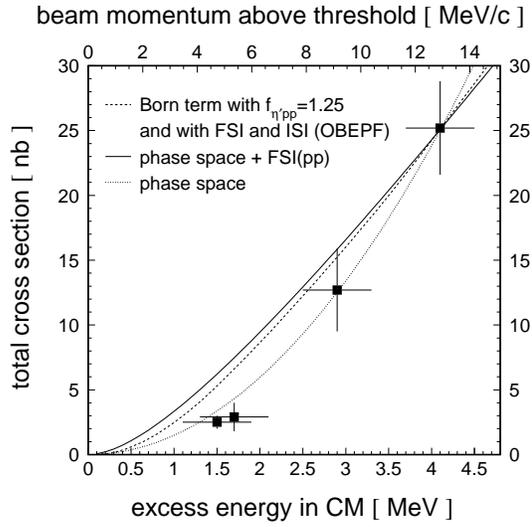}
   \vspace{3cm}
\caption{ 
Total cross sections for the $pp \rightarrow pp\eta^{\prime} $ reaction as a
function of the excess energy (bottom horizontal axis) and beam momentum above the threshold at
3.208 MeV/c (upper horizontal axis). 
The different lines show estimates for cross sections as described in the figure and outlined in
the text, where the curves are normalized to the data point at 4.1 MeV.
}
    \label{figure2.3}      
\end{figure}

\begin{center}
 P.~Moskal$^{1, 2}$,
 J.T.~Balewski$^{3, 2}$,
 A.~Budzanowski$^3$,
 H.~Dombrowski$^4$,
 C.~Goodman$^5$,
 D.~Grzonka$^2$,
 J.~Haidenbauer$^2$,
 C.~Hanhart$^2$, 
 L.~Jarczyk$^1$,
 M.~Jochmann$^6$,
 A.~Khoukaz$^4$,
 K.~Kilian$^2$,
 M.~K\"ohler$^6$,
 A.~Kozela$^3$,
 T.~Lister$^4$,
 U.-G.~Mei\ss{}ner$^2$,
 N.~Nikolaev$^2$,
 W.~Oelert$^2$,
 C.~Quentmeier$^4$,
 R.~Santo$^4$,
 G.~Schepers$^4$,
 U.~Seddik$^7$,
 T.~Sefzick$^2$,
 J.~Smyrski$^1$,
 M.~Soko\l owski$^1$,
 A.~Strza\l kowski$^1$, 
 C.~Thomas$^4$,
 M.~Wolke$^2$,
 P.~W\"ustner$^6$,
 D.~Wyrwa$^{1, 2}$
\end{center}
\noindent
$^1$ Institute of Physics, Jagellonian University, Cracow, Poland \\
$^2$ IKP,  Forschungszentrum J\"ulich,  Germany \\
$^3$ Institute of Nuclear Physics,  Cracow, Poland \\
$^4$ IKP, Westf\"alische Wilhelms--Universit\"at, M\"unster, Germany \\
$^5$ IUCF, Bloomington, Indiana, USA \\
$^6$ ZEL,  Forschungszentrum J\"ulich,  Germany \\
$^7$ NRC, Atomic Energy Authority,  Cairo, Egypt \\
\begin{flushleft}
The $pp \rightarrow pp \eta^{\prime}$ (958) reaction has been measured at COSY 
using the internal beam and the COSY-11 facility.  
The total cross sections at the four different excess energies
\mbox{$ Q = ~1.5 ~MeV, ~1.7 ~MeV, ~2.9 ~MeV,$ and $ ~4.1~MeV$} have been evaluated to be
\mbox{$ \sigma = 2.5 \pm 0.5~nb$, $~~~ 2.9 \pm 1.1~nb$, $~~~ 12.7 \pm 3.2~nb$, ~ 
and $~~~ 25.2 \pm 3.6 ~nb $},  respectively. \\
In this region of excess energy the $\eta^{\prime}$ (958) cross sections are much lower 
compared to those of the $\pi ^0$ and $\eta$ production.

\end{flushleft}
\newpage
The first experimental evidence of the $\eta^{\prime}$ meson has been 
seen in the \mbox{$K^-+ p \rightarrow \Lambda ^0 + neutrals$} reaction channels in 1964 
\cite{Kalb, Gold}.
Nowadays, the $\eta^{\prime}$(958) is well established as the heaviest member of the 
ground state pseudoscalar meson nonet with quantum numbers $I^G (J^{PC})$ = $0^+(0^{-+})$. 
The physics of the $\eta^{\prime}$ meson is related to one of the most intricate phenomena 
in particle physics.
In quark models \cite{Glashow} a nearly massless flavour singlet partner $\eta^{\prime}$
to the well established octet of pseudoscalar Goldstone bosons must exist.
With the advent of quantum chromo dynamics (QCD), however, the situation changed 
dramatically and there is no necessity~\cite{FGML} for a massless $\eta^{\prime}$. 
Without this U(1) anomaly~\cite{Weinb}, the $\eta^{\prime}$ would be 
unacceptably light: \mbox{$m_{\eta^{\prime}}^2 \leq 3m_{\pi}^2$}. Consequently, 
t'~Hooft~\cite{GHooft} has stimulated an extensive dispute on how the U(1) anomaly and QCD 
instantons affect the mass spectrum of the \mbox{$J^P$ = $0^-$} mesons \cite{U1} - \cite{Kura}.
The issues of ~~\mbox{i)~$\eta$ -- $\eta^{\prime}$} mixing,~~ii)~possible non-quarkonic component 
within the $\eta^{\prime}$ meson, and~~iii)~coupling of the $\eta^{\prime}$ to gluons have 
attracted much attention but the situation is far from being 
settled\mbox{~\cite{Diek} - \cite{Jous}}. Recently the CLEO \cite{CLEO} collaboration reported 
an anomalously large branching ratio for the inclusive decay of beauty particles 
$B \rightarrow \eta^{\prime} + X$, which is vitally discussed as evidence for strong coupling 
of $\eta^{\prime}$ meson to gluonic components \cite{Hutt} - \cite{Kagan}.\\
\\
There is no direct experimental information on the strength of the  $\eta^{\prime}$ 
coupling to nucleons: $g_{\eta^{\prime} NN}$. The smallness of the SU(3) singlet axial
charge current extracted from deep inelastic scattering data suggests a small 
$\eta^{\prime} NN$ coupling constant~\cite{TGV}.
On the other hand, the $\eta^{\prime}$ - nucleon coupling constant $g_{\eta^{\prime} NN}$ 
can put constraints on the theoretical quark models \cite{Zha, Dum}. 
Because there are no known "doorway like" $N\eta^{\prime}$ resonances close to the production
threshold, measurements of the cross sections for the $pp \rightarrow pp \eta^{\prime}$ reaction 
at such energies give an opportunity to determine the value of $g_{\eta^{\prime} NN}$. 
In case of the $\eta$ production, however, a reaction mechanism mediated by the intermediate 
resonance $N^*$ ($S_{11} (1535)$) is known to be important~\cite{CoWi, Vett} making an extraction
of the $\eta$ - nucleon coupling constant $g_{\eta NN}$ very difficult.  \\
\\
Recently data were published concerning the $\eta^{\prime}$(958) meson production in the 
$pd \rightarrow {}^3He~X$ reaction performed at SATURNE using the SPES4 spectrometer
\cite{Wur}. Assuming a pure s-wave phase space distribution the measured differential
cross section $d\sigma_{\eta^{\prime}} / d \Omega^* ~=~ 13 ~pb/sr $ results in a total 
cross section of $\sigma_{\eta^{\prime}} \approx ~ 0.16 ~nb $ at a mean excess energy of 
$Q = 0.5~MeV$. No data are published concerning the production of $\eta^{\prime}$ at
threshold in proton-proton collisions. There are only preliminary results from measurements at 
SATURNE \cite{Wilp}. Thus, the $\eta^{\prime}$ is the last non-strange meson of the 
pseudoscalar nonet for which cross sections for the production in 
the elementary proton proton scattering are unknown close to threshold.\\
\\
Measurements of the $\eta^{\prime}$ production in the pp interaction were performed at the
cooler synchrotron COSY-J\"ulich \cite{Bech} using an internal cluster target \cite{tar} 
in front of a regular C-shaped COSY dipole magnet acting as a magnetic spectrometer.
The $\eta^{\prime}$ mesons were not identified directly but their four-momentum vectors 
were determined via the missing mass method. The two outgoing protons
were registered in a set of two drift chamber stacks followed by a scintillator
hodoscope arrangement and a large area scintillator wall placed nine meter downstream.
Tracing the proton tracks back through the known three dimensional magnetic field
into the target spot results in a definite momentum determination. With the measured time 
of flight a unique particle identification is possible and therefore the four momentum 
vector components are given.  Details of the experimental apparatus are given elsewhere 
\cite{Brauk}. Measurements were performed at constant proton beam momenta as well as 
during a continuous beam momentum increase corresponding to excess energies from 
$Q~=~-3~MeV$ to $Q =~+5~MeV$. The total cross sections for four different excess energies:
\mbox {$Q~=~1.5~MeV,~1.7~MeV,~2.9~MeV,~$and$~4.1~MeV $} were evaluated. 
Figure \ref{eta_prime_missing_mass}a compares the experimental yield of the reaction 
$pp \rightarrow ppX $ measured just below the $\eta^{\prime}$ production threshold 
(solid line) to a phase space Monte Carlo (MC) calculation for the two and three pion 
production (dashed line). The broad structureless shape is well reproduced and thus 
explains the background. At the present value of the beam momentum up to seven 
pions could be produced in the pp scattering, however, due to the decreasing cross section 
with increasing number of pions these reactions do not contribute significantly. \\
Figure \ref{eta_prime_missing_mass}b shows the same experimental yield of the 
$pp \rightarrow ppX $ measurement
below the $\eta^{\prime}$ threshold (solid line) compared to the smoothed representation 
(dashed line) of these data which is used in the following to determine the reaction yield 
of the $\eta^{\prime}$ production above the unavoidable background. 
A small difference in shape between the two determinations of the background - the MC 
calculations and the smoothed sub-threshold measurement - is obvious. The $\eta^{\prime}$
yield evaluated by using the smoothed sub-threshold measurement as the background is 
systematically $(7 \pm 2) \% $ larger than applying the MC method. For the further 
analysis the experimentally determined smoothed sub-threshold background subtraction was 
used. 
In Fig.~\ref{eta_prime_missing_mass}c (similar as in Fig.~\ref{eta_prime_missing_mass}a and 
Fig.~\ref{eta_prime_missing_mass}b) the kinematical upper missing mass limit for the below 
threshold measurement is calibrated to the one above threshold. The clear $\eta^{\prime}$ peak 
is even more evident when subtracting both reaction yields from each other (above threshold 
minus below threshold) after normalization to the integrated luminosity, as seen in 
Fig.~\ref{eta_prime_missing_mass}d. The small seemingly structure at missing mass values below 
the $\eta^{\prime}$ mass is not significant from statistical point of view and since it does 
not reproduce itself for measurements at the other beam momenta. 
The counting rates have been corrected by extensive MC calculations for the detector 
acceptance and reconstruction efficiency, where the geometrical detector acceptance drops from
100 \% at threshold to 17 \% at \mbox {$Q~=~4.1~MeV $ }. For the detector acceptance $ E_{ff}$ 
the p-p final state interaction and the Coulomb repulsion were taken into account as outlined
in ref.~\cite{Bal}.\\
Simultaneously to the reaction under investigation elastically scattered protons 
have been recorded on tape and analysed. The differential cross section in the 
angular range of $\cos \Theta_{CM} ~=~ 0.45~$to$~0.75 $ was 
extracted and normalized to the EDDA data \cite{EDDA}, in order to determine the 
luminosity which varied during the running periodes between 
$ l = 4 \times 10^{29} cm^{-2}s^{-1} $ and $ l = 8 \times 10^{29} cm^{-2}s^{-1} $. 
Denoting the integrated luminosity by L and the entries in the $\eta^{\prime}$ peak by N,
the energy dependent total cross sections were evaluated according to:
${\sigma(Q) = N/(L \times E_{ff}(Q))}$.\\
The absolute beam momentum was calculated from the position of the $\eta^{\prime}$ peak in 
the missing mass spectrum. 
The spread in the beam momentum has been controlled by the sum signal of a beam position 
monitor from a longitudinal Schottky scan \cite{Bech} to be $\Delta p = 1.1~MeV/c$.
The inaccuracy of the missing mass evaluation originates besides 
from the beam momentum inaccuracy itself from the
uncertainty in the computation of the four-momentum vectors of the registered two protons.
That, in turn, can be caused by i) a misalignment of the angles of the drift chambers 
relative to the chosen coordinate system, ii) an uncertainty in the definition of the 
interaction point in both vertical and longitudinal directions, and iii) the inaccuracy 
of the knowledge of the dipole magnetic field. All these possible sources of miscalibration
were carefully studied by means of the COSY-11 MC program. It was established \cite{Mos1}
that these effects result in an error on the reconstructed missing mass of less than 0.4 MeV  
corresponding to an uncertainty in the absolute beam momentum of 1.2 MeV/c.\\
Figure \ref{figure2.3} depicts the values of the total cross section.   
The vertical error bars denote the statistical errors only.
The overall systematical error amounts to 15 \%, where 10 \% comes from the determination
of the detection efficiency $E_{ff}$ and 5 \% from the luminosity calculation. The horizontal 
error bars result from the inaccuracy of the absolute beam momentum determination \cite{Mos1}.\\

\begin{figure}[ht]
 \leavevmode
 \epsfxsize=8.5cm
   \epsfysize=16.5cm
   \hspace{1.5cm}
   \epsffile{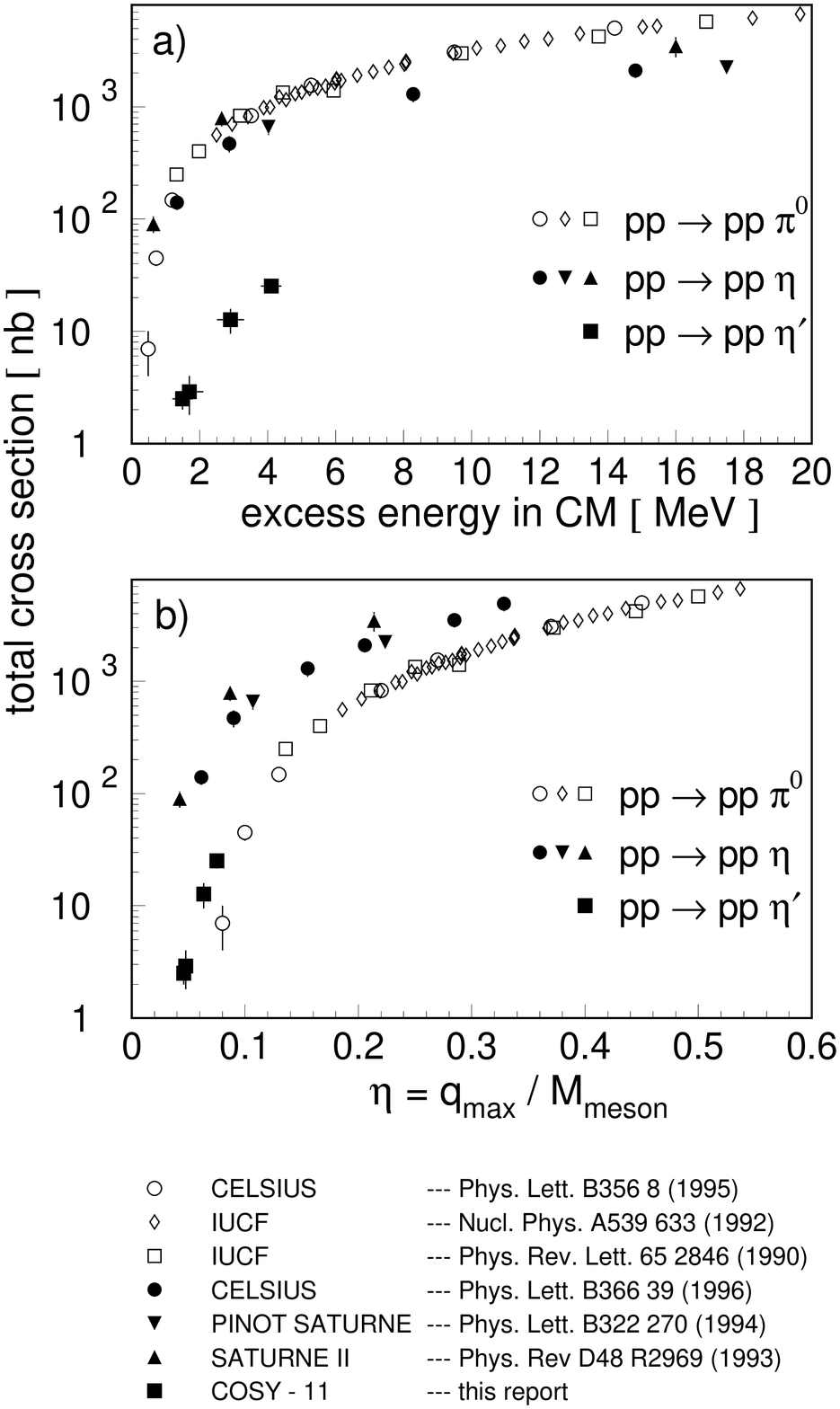}
   \vspace{-2.0cm}
\caption{Total cross sections for the reactions: 
$pp \rightarrow pp \pi^0,~~pp \rightarrow pp \eta $, and
$~~pp ~\rightarrow pp \eta^{\prime}$ \protect \\
~~~~~~a) as a function of the excess energy and \protect \\
~~~~~~b) as a function of the maximum meson momentum normalized to the meson mass.
} 
\label{figure_neu_2}
\end{figure}

In Fig.~\ref{figure_neu_2} a comparison of the 
$pp \rightarrow pp \pi^{0},~pp \rightarrow pp \eta,~and~
pp \rightarrow pp \eta^{\prime}$ total cross sections is presented.
Figure~\ref{figure_neu_2}a depicts the production cross sections as a function of the respective
excess energy,
where we observe that the cross section ratio for the $\pi^{0}$/$\eta^{\prime}$ production scales 
approximately with the square of the mass ratio $(135/958)^2~\approx~0.02 $, indicating a similar 
production process. Here the $\eta$ production cross section is, however, much larger which can be 
attributed to a dominant contribution of the $S_{11}(1535)$ resonance. In fact, on this scale
the two mesons $\eta$ and $\pi^{0}$ are produced with rather similar cross sections, whereas the 
reaction yield for the $\eta^{\prime}$ is more than one order of magnitude smaller, see also 
ref. \cite{Bornec}.\\
Representing the total cross sections as a function of the $\eta$ variable,
where the parameter $\eta$ stands for the maximum center of mass meson momentum normalized to 
its mass, the $pp \rightarrow pp \eta^{\prime}$ reaction yield is similar to the one of the 
$pp \rightarrow pp \pi^0$ data in contrast to the much larger $\eta$ meson production rate,
as is shown in Fig.~\ref{figure_neu_2}b. This again suggests that the production mechanisms for 
$\pi^{0}$  and $\eta^{\prime}$ are similar.\\
The theory of $\eta^{\prime}$ production is in its formative stage. Whereas in the case of the 
$\eta$ meson the production via the $S_{11} (1535)$ resonance is dominant~\cite{CoWi, Vett} there 
are no obvious candidates for baryon resonances decaying into $\eta^{\prime} (958)$ and the nucleon, 
apart from the  $D_{13} (2080) $ resonance \cite{Zha} which, due to its spin s = 3/2, should have 
only a very suppressed influence on the reaction process at threshold. Therefore, as a first 
approximation, one can consider the effective Lagrangian approach with direct $\eta^{\prime}$NN 
coupling (for a related discussion of photoproduction see ref.~\cite{Zha, Benm}). Alongside with 
i) the pure phase space distribution (dotted line) and ii) the phase space distribution including the 
pp final state interaction \cite{Haidenbauer} (solid line) (which is known to be important 
\cite{Bond, Calen}, and calculated as outlined in ref~\cite{Bal}), the result of such a model 
evaluation is shown in Fig.~\ref{figure2.3} by the dashed line. The disagreement between the energy 
dependence obtained under these simple assumptions with the experimental data indicates that heavy 
meson exchange or other mechanisms may contribute significantly to the production of the 
$\eta^{\prime}$ meson in the $pp \rightarrow pp \eta^{\prime}$ reaction. \\
With the assumption that the production of the $\eta^{\prime}$ meson is driven by the direct 
term only and that the production amplitude from the heavy meson exchange has the same sign as the 
amplitude of the direct term \cite{Hanhart} the upper limit for the coupling constant can be estimated. 
By normalizing the theoretical result to the data point at $Q~=~4.1~MeV$ the pseudoscalar coupling 
constant $g_{\eta^{\prime}pp}$ turns out to be smaller than 2.5, where 
predictions \cite{Zha, Dum, Efr} for $g_{\eta^{\prime}pp}$ range from values 1.9~to~7.5
and the dispersion method \cite{Kroll} gives  $g_{\eta^{\prime}pp}$ values consistent with 
zero.\\

In short, evidence has been given by the present studies of the  
$pp \rightarrow pp\eta^{\prime}$ reaction at threshold that i) there seems to be
no indication that an S - wave ($N \eta^{\prime}$) $N^*$ resonance intermediate 
"doorway like" state governs the reaction mechanism and that 
ii) the $\eta^{\prime}$coupling constant
$g_{\eta^{\prime}pp}$ extracted from a simple model analysis appears to be consitent with
the range expected by the quark model, barring an accidental cancellation between 
interferences of the amplitudes for the direct term and the heavy meson exchange. \\
\\
\\
\\
\\
We appreciate the work provided by the COSY operation team and we thank them for the good 
cooperation and for delivering the excellent proton beam.
\\
We would like to thank our COSY-11 adviser C.~Wilkin for inspiring and helpful discussions.
\\
The research project was supported by the BMBF, the Polish Committee for 
Scientific Research, and the Bilateral Cooperation between Germany and Poland
represented by the Internationales B\"uro DLR for the BMBF. The collaboration partners
from both the Westf\"alische Wilhelms-University of M\"unster and the Jagellonian
University of Cracow appreciate the support provided by the FFE-grant from the 
Forschungszentrum J\"ulich.
 
\newpage

\newpage

\newpage

\end{document}